\documentclass[aps,prb,twocolumn,superscriptaddress,showpacs]{revtex4}

\usepackage{graphicx}
\usepackage{dcolumn} 
\usepackage{bm}      
\usepackage{color}
\usepackage{float}
\newcommand{\ep}{$e$-$ph$~}
\newcommand{\pdag}{{\phantom\dagger}}
\newcommand{\bk}{{\bf k}}
\newcommand{\bq}{{\bf q}}

\begin{document} 

\title{Properties of the non-linear Holstein polaron at finite doping and temperature}
\author{Shaozhi Li}
\affiliation{Department of Physics and Astronomy, University of Tennessee, Knoxville, Tennessee 37996-1200, USA}

\author{E. A. Nowadnick}
\affiliation{School of Applied and Engineering Physics, Cornell University, Ithaca, NY 14853 USA}

\author{S. Johnston}
\affiliation{Department of Physics and Astronomy, University of Tennessee, Knoxville, Tennessee 37996-1200, USA}

\date{\today}
 
\begin{abstract}
We use determinant quantum Monte Carlo to  
study the single particle properties of quasiparticles and phonons  
in a variant of the two-dimensional Holstein model that includes an additional 
non-linear electron-phonon (\ep) interaction.  
We find that a small positive non-linear interaction reduces the 
effective coupling between the electrons and the lattice, 
suppresses charge-density wave (CDW) correlations, and hardens the 
effective phonon frequency. Conversely, 
a small negative non-linear interaction can enhance the \ep coupling 
resulting in heavier quasiparticles, an increased tendency towards 
a CDW phase at all fillings, and a softened 
phonon frequency. An effective
linear model with a renormalized interaction strength and phonon frequency 
can qualitatively capture this physics; 
however, the quantitative effects of the non-linearity on both the 
electronic and phononic degrees of freedom cannot be captured by 
such a model. These results are significant for typical non-linear 
coupling strengths found in real materials, indicating that non-linearity 
can have a significant influence on the physics of many \ep 
coupled systems. 
\end{abstract}


\maketitle 

\section{Introduction}
The electron-phonon ($e$-$ph$) interaction plays an important role in many systems
including conventional metals and superconductors, \cite{Gruner, Parks} 
organic semiconductors,\cite{Nenad, Shaozhi} fullerenes,\cite{fullerene1,fullerene2} 
and a large number of transition metal oxides.\cite{Lanzara,tpd,LeePRL2013,King, Park, 
MedardePRL1998,ParkPRL2012,JohnstonPRL2014, Jaramillo,MannellaNature2005,LevPRL2015}  
In general, the \ep interaction induces a local distortion of the 
lattice surrounding a carrier, resulting in quasiparticles dressed by phonon 
excitations known 
as polarons. In some cases these lattice distortions can be large and tightly 
bound to the electron, such that the quasiparticle has a very large effective 
mass $m^*$ and vanishing quasiparticle residue $Z$.\cite{PolaronReview}  
In this limit, the quasiparticle is typically referred to as a small polaron. 
In most models, the crossover to 
the small polaron regime occurs for $\lambda > 1$,  
where $\lambda$ parameterizes the (dimensionless) strength of the \ep interaction.
When $\lambda < 1$ the carriers are still dressed by the lattice, 
forming large polarons where the lattice distortions are spread out over 
many lattice constants. 

Almost all of our knowledge about the effects of the \ep interaction 
has been obtained from linear models. The derivation of these models is standard. 
First, the \ep interaction is expanded in powers of the 
atomic displacement. This is followed by a  
truncation of the expansion to linear order under the {\it assumption} of small 
lattice displacements. These same models, however, often predict the formation of small polarons 
or charge-density-wave (CDW) phases for sufficiently large \ep coupling, 
which are characterized by large lattice displacements.  
For example, the displacements in the linear Holstein and Hubbard-Holstein 
models can be on the order of the lattice constant when CDW correlations are 
significant.\cite{JohnstonPRB2013,LiEPL2015,CraigPRB2007} This can occur 
even for weak values of the \ep coupling if the Fermi surface is well nested, as 
is the case for the Holstein model with nearest neighbor hopping on a cubic lattice.  
Similarly, a small polaron can be dressed
by tens to hundreds of phonon quanta,\cite{MA} implying the presence of a
heavily distorted lattice surrounding the carrier. Clearly, these predictions directly 
violate the assumptions underlying the linear model, which is an unambiguous sign 
that important physics has been discarded during its derivation.\cite{AdolphsEPL2013} 

These considerations show that the higher order terms in the \ep 
interaction are likely to be important
whenever the linear term is large (or  when strong nesting conditions are 
present). But non-linear and anharmonic
effects can also be ``switched on" in a weakly coupled system, if the underlying atoms of the 
lattice are driven far from their equilibrium positions by an external perturbation. 
For example, several recent 
experiments have exploited optical pump pulses to drive the lattice, creating
large lattice deformations or exciting coherent phonon oscillations.\cite{Pump1,Pump2,Pump3,Pump4}  
These excited 
lattice states can melt various ordered phases\cite{Melt1,Melt2} or can 
induce transient superconducting states.\cite{KaiserPRB2014,MankowskyNature2014} 
Here, nonlinear and anharmonic phonon dynamics are thought to play a vital 
role in creating such transient states.\cite{KaiserPRB2014,MankowskyNature2014,Pump2} 

The first 
attempts to include higher order terms were made by Adolphs and Berciu (Ref. \onlinecite{AdolphsEPL2013}). They   
considered the effects of non-linear \ep interactions on a single polaron in the 
Holstein model using the momentum average approximation\cite{MA} and found that
small non-linear couplings dramatically undress the quasiparticle. 
This was attributed to a simultaneous hardening of the bare phonon frequency and
a renormalization of the bare \ep coupling constant, resulting in an overall 
weaker effective linear interaction.  
Later work by some of the present authors considered 
finite carrier concentrations and temperatures using non-perturbative determinant 
quantum Monte Carlo (DQMC).\cite{LiEPL2015} Here too, the
presence of a non-linear interaction was found to suppress the tendency towards the
formation of CDW and superconducting states found in the linear model.     

\begin{figure}[t]
 \includegraphics[width=0.9\columnwidth]{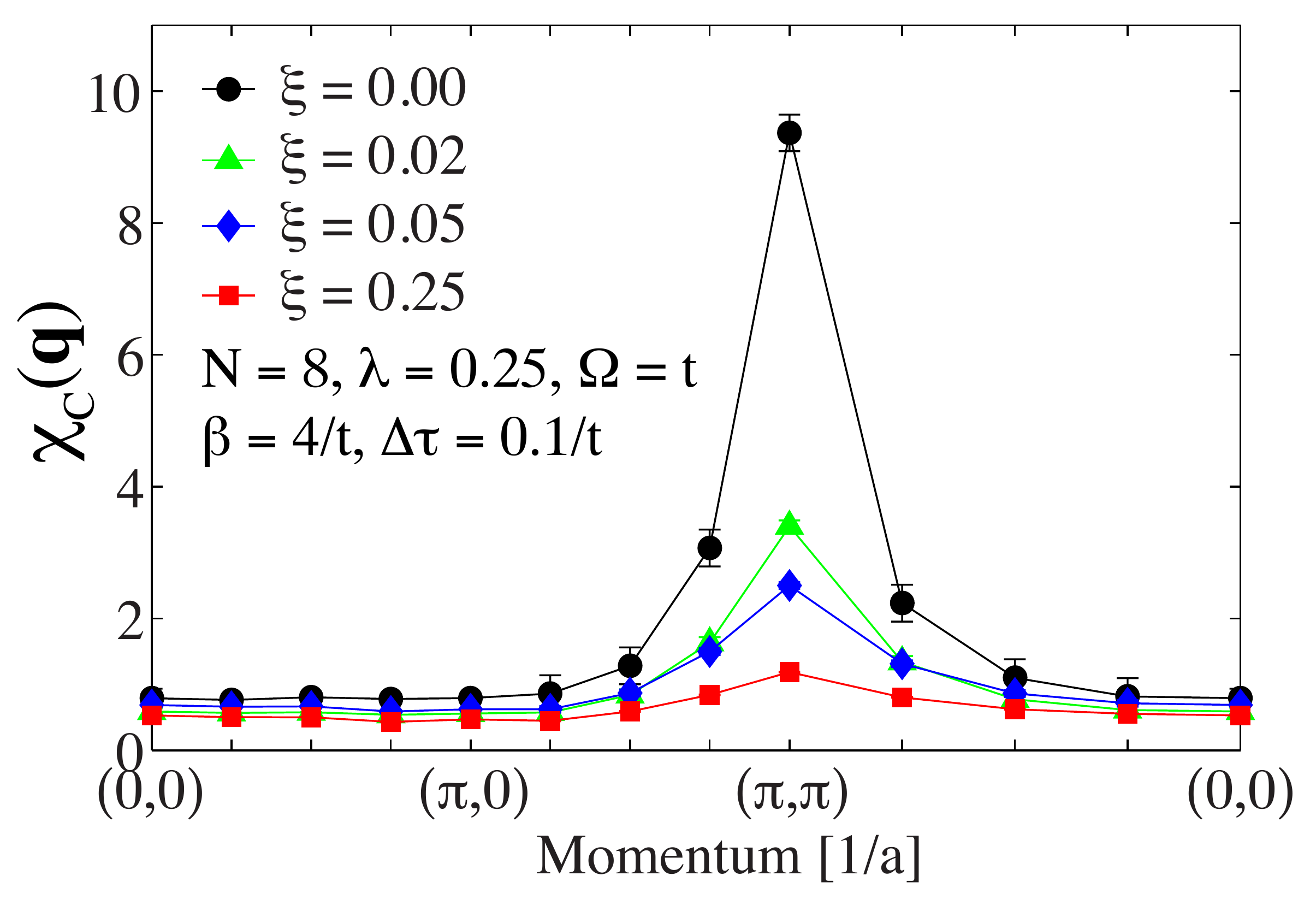}
 \caption{\label{Fig:Chiq} (color online) The momentum dependence of the charge 
 susceptibility $\chi_C(\bq)$ as a function of non-linear interaction strength $\xi > 0$. 
 The trends shown here follow those reported in Ref. \onlinecite{LiEPL2015}, where the 
 increasing non-linear interaction strength suppresses the CDW correlations in the system.  
 The parameters for the calculation are as indicated. 
 Error bars smaller than the marker size have been suppressed for clarity. 
 }
\end{figure} 

It is important to note that the effects of the non-linear interactions 
observed in Refs. \onlinecite{LiEPL2015} and \onlinecite{AdolphsEPL2013} are 
significant, even for relatively small non-linear interaction strengths.  
This is illustrated in Fig. \ref{Fig:Chiq}, which shows the suppression of the 
CDW correlations in the half-filled Holstein model as a function of the 
non-linear coupling strength. [The degree of non-linearity is
indicated by the ratio of the pre-factors of the quadratic ($g_2$) and
linear ($g_1$) interaction terms $\xi = \frac{g_2}{g_1}$, see Sec. \ref{Sec:Model}.] 
Here, the charge susceptibility provides a measure of the strength of the 
CDW correlations. It is defined as  
\begin{equation}
\chi_C(\bq) =  \frac{1}{N^2}\int_0^\beta d\tau \langle  
\hat{T}_\tau\hat{\rho}(\bq,\tau)\hat{\rho}^\dagger(\bq,0)\rangle , 
\end{equation}
where $\hat{\rho}(\bq) = \sum_{i,\sigma} e^{i\bq\cdot {\bf R}_i}\hat{n}_{i,\sigma}$, 
and $\hat{T}_\tau$ is the time-ordering operator.  
In this case, the well-known\cite{MarsiglioPRB1990,ScalettarPRB1989} 
tendency for the linear model to form  
a ${\bf Q} = (\pi/a,\pi/a)$ CDW at half-filling 
is suppressed for $\xi$ as small as $\sim 0.02 - 0.05$. 

The suppression of $\chi_C(\bq)$ for small $\xi$ 
is noteworthy because this ratio is typical of many models 
used to parameterize \ep interactions in real materials. In the 
transition metal oxides, for example, electrons can couple strongly to  
oxygen bond stretching modes via the modulation of the near-neighbor 
transition metal 3$d$-oxygen 2$p$ (TM-O) hopping integral $t_{pd}$, 
which depends on the TM-O bond distance.\cite{Rosch,JohnstonPRL2014} 
If $d$ is TM-O bond distance and $d_0$ is its equilibrium value, 
then this coupling mechanism leads to terms in the Hamiltonian of the form  
\begin{equation}\nonumber 
H_\mathrm{kin} = \sum_{\langle i,j\rangle,\sigma} t_{pd}(d) c^\dagger_{i,\sigma}p^\pdag_{j,\sigma},
\end{equation}
where the $c_{i,\sigma}$ ($p_{j,\sigma}$) operators act on the TM 3d and O 2p orbitals, 
respectively. The hopping integral has a typical dependence  
$t_{pd} \sim \left(d/d_0\right)^{-\beta}$, with $\beta = 3.5$.\cite{Rosch} 
Expanding this dependence to second order in powers of $(d-d_0)$ gives 
\begin{equation}\nonumber
H_\mathrm{e-ph} = \sum_{\langle i,j\rangle,\sigma} \left[\alpha_1 (d-d_0) + 
\alpha_2 (d-d_0)^2\right] c^\dagger_{i,\sigma}p^\pdag_{j,\sigma}, 
\end{equation}
where $\alpha_1 = -\beta t_{pd}(d_0)/d_0$ and 
$\alpha_2 = \frac{\beta(\beta+1)t_{pd}(d_0)}{2d_0^2}$. (The zeroth order terms 
are included in the non-interacting terms of the Hamiltonian). Taking typical 
values\cite{Footnote} for the various parameters in transition metal oxides, 
one arrives at a ratio of $|\xi| \sim 0.05$. Therefore real materials have intrinsic 
non-linear interactions that are large enough to be relevant when the linear coupling 
is strong. For this particular coupling mechanism the sign of $\xi$ is negative, 
however, this is not guaranteed for all coupling mechanisms.\cite{AdolphsPRB2014}  

In this paper we expand upon our previous work examining the impact of non-linear 
interactions on the CDW and superconducting phases of the Holstien model.\cite{LiEPL2015} 
We present results for the single-particle electronic and phononic 
properties of the model, thus providing a more comprehensive picture of the 
effects of non-linearity.
Since DQMC is formulated in the grand canonical ensemble, we are able to examine 
these properties at finite temperatures and carrier concentrations for the first time. 
We find that 
the inclusion of a non-linear interaction renormalizes both the effective frequency 
of the Holstein phonon and the effective \ep coupling strength, resulting in significant  
changes in both the electronic and phononic properties of the model.  
Furthermore, we demonstrate that while the qualitative effects of the non-linearity 
can be framed in terms of an effective linear model, the quantitative effects 
on both the electronic and phononic properties of the model cannot be. 
This conclusion requires an examination of both the electronic and phononic properties, 
and thus cannot be arrived at by considering electronic properties only. 
Our results reenforce the notion that the full non-linearity must be included 
in order to obtain an accurate picture of both the 
electronic and phononic degrees of freedom whenever strong 
linear \ep interactions are present. 

\section{Methods}\label{Sec:Methods}
\subsection{The Non-linear Holstein Model}\label{Sec:Model}
We study a variant of the Holstein model that includes additional non-linear 
interaction terms. The Hamiltonian is partitioned as  
\begin{equation}\label{Eq:Full}
H = H_\mathrm{el} + H_\mathrm{lat} + H_\mathrm{int},  
\end{equation}
where  
\begin{equation}
H_\mathrm{el} = -t\sum_{\langle i,j\rangle,\sigma} 
c^\dagger_{i,\sigma}c^{\phantom\dagger}_{j,\sigma} - \mu \sum_{i,\sigma}\hat{n}_{i,\sigma}, 
\end{equation} 
contains the non-interacting electronic terms, 
\begin{equation}
H_\mathrm{lat}=\sum_{i} \left[\frac{\hat{P}_i^2}{2M} + \frac{M\Omega}{2}\hat{X}_i^2 \right] 
= \sum_i \Omega \left[b^\dagger_ib^{\phantom\dagger}_i+\frac{1}{2}\right],  
\end{equation}
contains the non-interacting lattice terms, 
and 
\begin{equation}
H_\mathrm{int}=\sum_{i,k,\sigma} \alpha_k \hat{n}_{i,\sigma}\hat{X}_i^k =\sum_{i,k,\sigma} 
g_k \hat{n}_{i,\sigma}(b^\dagger_i + b^{\phantom\dagger}_i)^k 
\end{equation}
contains the interaction terms to $k$\textsuperscript{th} order in the 
atomic displacement. Here, $c^\dagger_{i,\sigma}$ ($c^{\phantom\dagger}_{i,\sigma}$) 
creates (annihilates) an electron of spin $\sigma$ on lattice site $i$; 
$b^\dagger_i$ ($b^{\phantom\dagger}_i$) creates (annihilates) a phonon 
on lattice site $i$; 
$\hat{n}^{\phantom\dagger}_{i,\sigma} = c^\dagger_{i,\sigma}c^{\phantom\dagger}_{i,\sigma}$ is 
the number operator; $\mu$ is the chemical potential; 
$t$ is the nearest-neighbor hopping integral; 
$M$ is the ion mass; $\Omega$ is the phonon frequency; $\hat{X}_i$ and 
$\hat{P}_i$ are the lattice position and momentum operators, respectively; 
and $g_k = \alpha_k(2M\Omega)^{-\frac{k}{2}}$ is the strength of the \ep  
coupling to $k$\textsuperscript{th} order in displacement. 

The non-linear Holstein model is characterized by several dimensionless 
parameters, and the specific choice in parameterization is not unique.
Here, we follow the convention used in previous works,\cite{AdolphsEPL2013,LiEPL2015} where 
the usual dimensionless 
parameter $\lambda = \alpha^2_1/(M\Omega^2W) = g_1^2/(4t\Omega)$ 
parameterizes the linear coupling strength 
and $\xi_k = g_k/g_{k-1}$ parameterizes the non-linear interaction terms.
This choice provides a convenient interpretation with 
large $\lambda$ implying a strong linear interaction 
and large $\xi_k$ implying strong non-linear effects.  
In the linear model ($\xi_k = 0$) $\lambda > 1$ implies the formation of 
small polarons. Thus this choice of parameterization is also useful for 
making comparisons to our expectations gained from studying the linear model.  

\subsection{Determinant Quantum Monte Carlo}
We use DQMC to study the  
non-linear Holstein model. The details of the method are given in 
several references (see for example 
Refs. \onlinecite{BSS} - \onlinecite{Chang})  
and the specifics for handling the lattice degrees of freedom can be found in Refs. 
\onlinecite{JohnstonPRB2013}, \onlinecite{LiEPL2015}, and \onlinecite{ScalettarPRB1989}.

In our calculations  
we keep $g_1>0$ without loss of generality. 
Furthermore, Ref. \onlinecite{AdolphsEPL2013} examined terms to 4\textsuperscript{th} 
order in the interaction and found that the largest effect was produced by 
the 2\textsuperscript{nd} order terms. We expect a similar result here and 
restrict ourselves to $k = 2$ while defining $\xi = \frac{g_2}{g_1}$. 
Throughout this work we examine two-dimensional square lattices  
with a linear dimension $N$ (a total of $N\times N$ sites) and set 
$a = t = M = 1$ as our units of distance, energy, and mass, respectively.    
We typically work on lattice sizes ranging from $N = 4$ to $8$ in size. 
In general we do not observe significant finite size effects,\cite{LiEPL2015}  
which is likely due to the local nature of the interaction in the model.  

The Holstein model and its non-linear extension do not 
suffer from a fermion sign problem.\cite{WhitePRB1989,SignProblem}
We are therefore able to perform simulations to arbitrarily low temperatures,  
\cite{LiEPL2015} however, we find that most of the physical properties we 
are interested in here can be examined for $\beta = 4/t$. 
We use this temperature for all plots in this work unless stated otherwise and 
present results for an imaginary time discretization of $\Delta\tau = 0.1/t$. 
In all of our simulations we have not observed any significant 
$\Delta\tau$ errors introduced by this choice.

\subsection{Analytic Continuation}
The DQMC calculation provides the phonon Green's function 
$D({\bf q},\tau) = \langle T\hat{X}_\bq(\tau)\hat{X}_{-\bq}(0)\rangle$ measured on 
the imaginary time axis. The phonon Green's function is related to the phonon spectral 
function on the real axis by the integral equation 
\begin{equation}
\int_0^\beta d\tau D(\bq,\tau) = \int_{-\infty}^\infty \frac{d\omega}{2\pi}
\frac{B(\bq,\omega)}{\omega}. 
\end{equation}
This equation also provides a normalization condition for $B(\bq,\omega)$. 
In section \ref{Sec:AC} we examine the phonon spectral properties of the non-linear 
Holstein model, which requires that the phonon Green's function be analytically 
continued to the real frequency axis. This is accomplished with the Maximum Entropy method  
(MEM).\cite{MEM} The analytic continuation procedure is identical to the one given in 
Ref. \onlinecite{NowadnickPRB2015} and the reader can refer to there for details.  

MEM requires a model default function to define the entropic prior.  
We adopt a momentum-independent Lorentzian model, which is peaked at the renormalized 
phonon frequency predicted by the mean-field treatment of the non-linear interaction 
$\Omega_{MF} = \Omega + 2g_2$ (see section \ref{Sec:MF}) and of width $t$. 
Our results are robust against reasonable changes in this choice of default function.  

\section{Results and discussion}\label{Sec:Results}
Our results for the single-particle electronic and 
phononic properties of the non-linear Holstein model are presented in this section. 
We will begin by first examining the renormalization of the quasiparticles 
as a function of doping, temperature, and phonon energy. We then  
examine how the interplay between the quasiparticles and renormalized 
phonons affect the energetics of the system. Following this, results are 
presented for the renormalization of the phonons. 
Finally, after all of these effects are examined, we demonstrate that the simultaneous 
renormalization of the electronic and phonic 
subsystems cannot be quantitatively captured by an effective linear model.  
Combined, these results paint a more detailed picture of how {\it both} the quasiparticles 
and phonons are renormalized by the non-linear \ep interaction, which 
cannot be obtained by examining only one of these subsystems.   
In the following sections we consider results for the case $\xi > 0$. Considerations 
of the $\xi < 0$ case are presented in Sec. \ref{Sec:xineg}. 

\subsection{The quasiparticle residue}
We begin by examining the polaron's quasiparticle reside $Z(\bk)$ 
as a function of the non-linear coupling strength and doping. This quantity 
is related to the effective mass via $Z^{-1} \propto \frac{m^*}{m}$. It  
can be obtained from the imaginary axis self-energy $\Sigma(\bk,i\omega_n)$ using 
the relationship $Z(\bk) = \frac{1}{1+b(\bk)}$, \cite{Arsenault}
where 
\begin{equation}\label{Eq:b}
b(\bk) = \lim_{\omega_n \rightarrow 0} -\frac{\partial \Sigma^\prime(\bk,i\omega_n)}{\partial \omega_n}.
\end{equation}
Here, we approximate $b(\bk)$ by evaluating Eq. (\ref{Eq:b}) for the lowest Matsubara 
frequency $\omega_n = \pi/\beta$. 

\begin{figure}
 \includegraphics[width=\columnwidth]{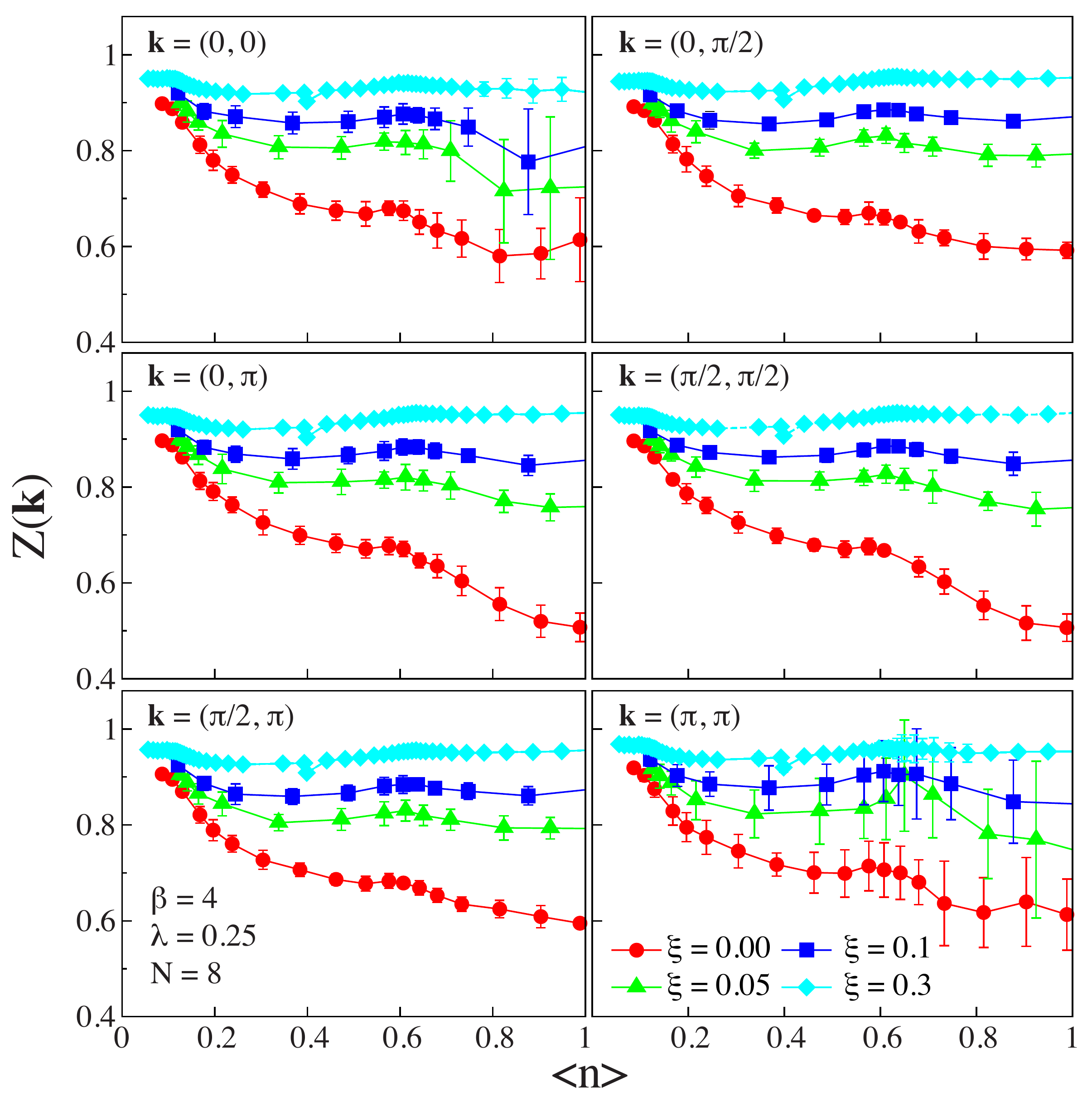}
 \caption{\label{Fig:Z_vs_n} (color online) 
 The quasiparticle residue $Z({\bf k})$ as a function of band 
 filling $\langle n \rangle$ for (a) ${\bf k} = (0,0)$, 
 (b) $(0,\pi/2)$, (c) $(0,\pi)$, (d) $(\pi/2,\pi2)$, 
 (e) $(\pi/2,\pi)$, and (f) $(\pi,\pi)$. 
 Results are shown for various values of the non-linear 
 interaction strength $\xi$, as indicated in panel (f), and 
 are obtained using an $N = 4\times 4$ cluster with a linear 
 coupling $\lambda = 0.25$ and an inverse temperature $\beta = 4/t$.   
 Error bars smaller than the marker size have been suppressed for clarity. 
 }
\end{figure}

Fig. \ref{Fig:Z_vs_n} shows $Z(\bk)$ as a function of carrier concentration 
for several values of the non-linear coupling $\xi$. These results were 
obtained on an $N = 4$ cluster, using a linear coupling strength $\lambda = 0.25$ 
and $\Omega = t$. In the linear model ($\xi = 0$, red dots)  
the quasiparticle residue decreases as the 
filling approaches $\langle n \rangle = 1$, where the 
${\bf Q} = (\pi,\pi)$ CDW correlations begin to dominate the system (Fig. \ref{Fig:Chiq}). 
Note that strong CDW correlations are observed, even for the small value of 
the linear coupling used here, due to a perfect $(\pi,\pi)$ nesting condition in the two-dimensional 
Fermi surface. This nesting condition also results in strong lattice displacements 
in the linear model. As a result, the inclusion of the non-linear terms has a 
significant effect on the quasiparticle residue where,    
for $\xi > 0$, a significant undressing of the 
polarons occurs and the quasiparticle residues at all momenta begin to rise. 
This occurs at all doping, however, the effect is more pronounced near half-filling.  
(Our $\xi = 0$ results are in good agreement with  
Ref. \onlinecite{MishchenkoPRL2014}, which examined larger system sizes using 
a complementary diagrammatic Monte Carlo method.)

\begin{figure}
 \includegraphics[width=0.7\columnwidth]{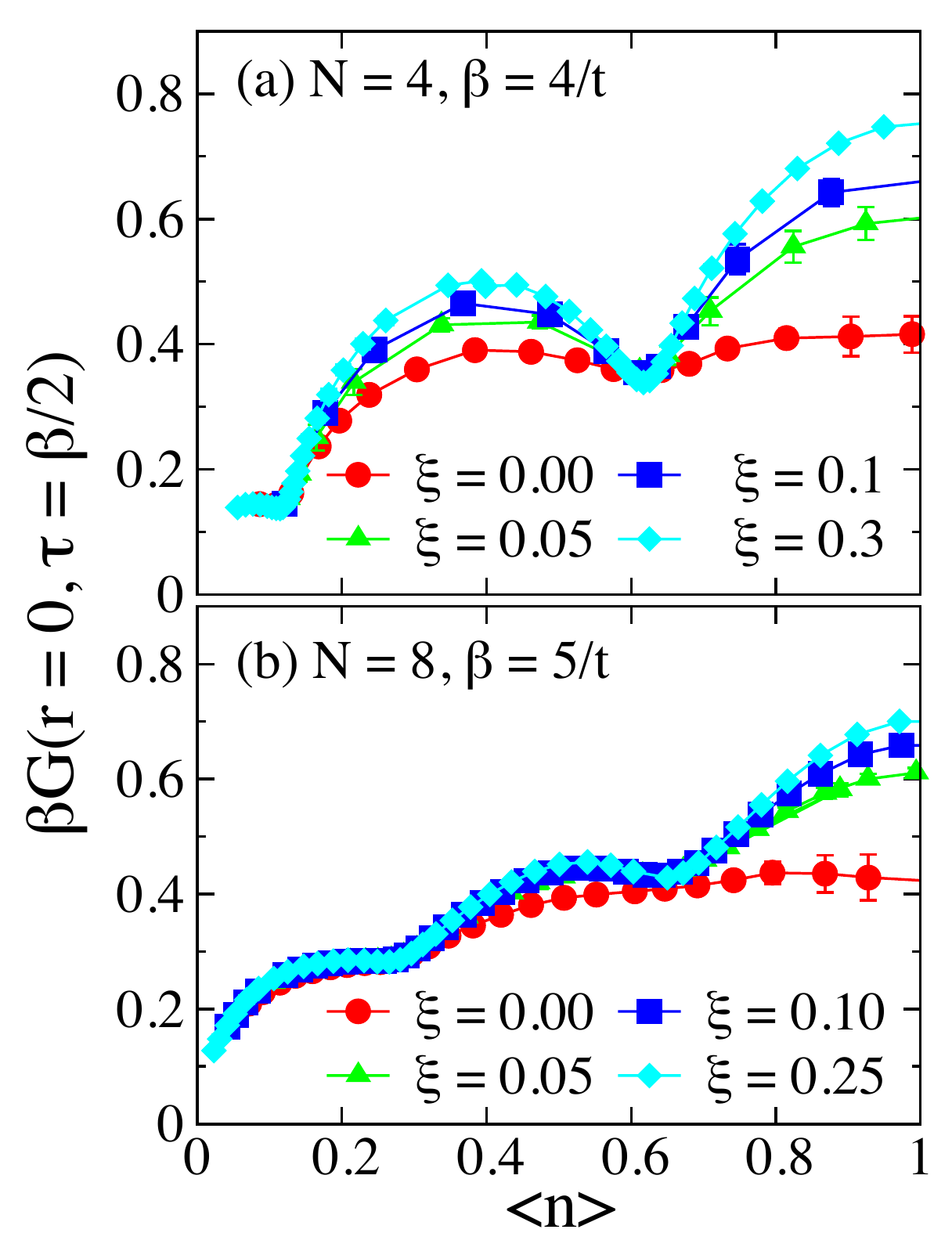}
 \caption{\label{Fig:spec_vs_n} The spectral weight at the Fermi level given by
 $\beta G(r=0,\tau=\beta/2)\equiv\beta G_{\beta/2}$ as a function of band
 filling $\langle n\rangle$ for various values of the non-linear coupling 
 strength $\xi$. 
 (a) Results for a $N = 4$ cluster and an inverse temperature $\beta = 4/t$. 
 (b) Results for a larger $N = 8$ cluster and a lower temperature $\beta = 5/t$. 
 All results are obtained for a linear coupling $\lambda = 0.25$ and a 
 frequency of $\Omega = t$. Error bars smaller than the markers
 have been suppressed for clarity.}
\end{figure} 

The formation and suppression of the CDW gap is also reflected in the spectral weight 
at the Fermi level, which can be obtained from the local imaginary time Green's function 
$\beta G(r=0,\tau=\beta/2)\equiv\beta G_{\beta/2}$.\cite{Trivedi} 
Fig. \ref{Fig:spec_vs_n}a plots $G_{\beta/2}$ as a function of  
filling $\langle n\rangle$ for the same parameters used in Fig.
\ref{Fig:Z_vs_n}. Fig. \ref{Fig:spec_vs_n}b plots similar results obtained on a 
larger cluster and at lower temperature, where the 
qualitative behavior is the same.  
The spectral weight in the linear model initially grows with
increasing carrier concentration, but saturates 
as the concentration approaches half-filling and CDW 
correlations begin to dominate. When a non-linear interaction is introduced, 
however, $G_{\beta/2}$ increases at most fillings, which 
is most pronounced near  $\langle n \rangle \sim 1$.  
(The dip around $\langle n \rangle=0.6$ is a finite size  
effect due to the smaller number of momentum points in the $N = 4$ cluster. 
It is much less pronounced on the larger $N = 8$ cluster.)  
This spectral weight increase directly reflects the increase in the 
quasiparticle residue and the suppression of the CDW correlations.  
Previously we showed that a large non-linear coupling drives
the system into a metallic state at half-filling, with the value of $\beta G_{\beta/2}$
approaching the non-interacting value.\cite{LiEPL2015} The results in Fig. \ref{Fig:spec_vs_n} indicate that this also occurs for carrier 
concentrations away from half-filling.

\begin{figure}
 \includegraphics[width=0.7\columnwidth]{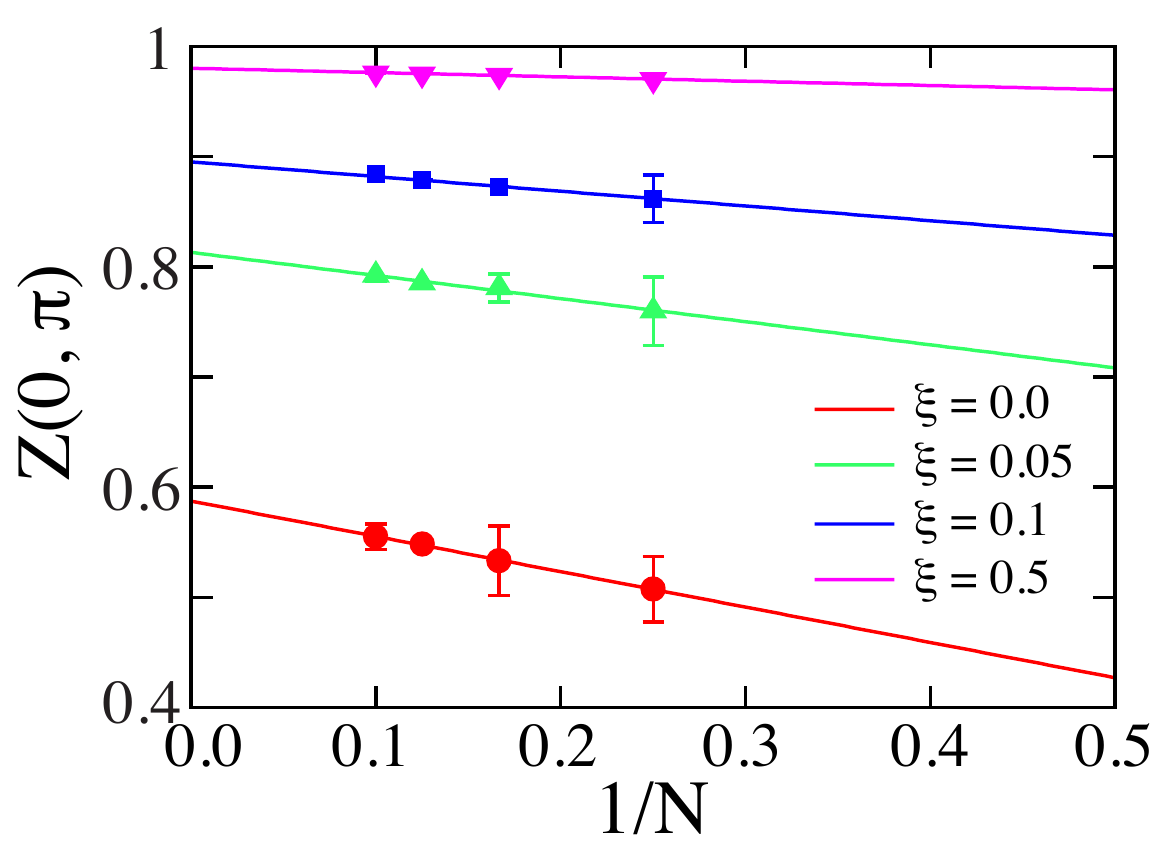}
 \caption{\label{Fig:Zscaling} (color online) A finite size scaling analysis of 
 $Z(0,\pi)$ as a function of $1/N$ where $N$ is the linear dimension of the 
 cluster. The parameters for the calculations are $\beta = 4/t$, $\Omega = t$, 
 and $\lambda = 0.25$. Error bars smaller than the marker size have been suppressed 
 for clarity. }
\end{figure} 

The results presented in Figs. \ref{Fig:Z_vs_n} and 
\ref{Fig:spec_vs_n}a are obtained on a $N = 4$ 
cluster; however, they are qualitatively representative 
of the results obtained for all examined cluster sizes, as hinted 
at by comparing Figs. \ref{Fig:spec_vs_n}a and \ref{Fig:spec_vs_n}b.  
To confirm this, in Fig. \ref{Fig:Zscaling} we 
perform a finite size scaling analysis for 
$Z(0,\pi)$ at half-filling, where the reduction in $Z$ by CDW correlations 
is most pronounced. From this analysis it is clear that the qualitative behavior 
is not affected by finite size effects and survives in the 
thermodynamic limit. Moreover, the more pronounced finite size effects occur 
when the non-linear interaction is weak. 
As similar scaling results were obtained 
for both the charge susceptibility and the electron spectral weight 
in our previous work,\cite{LiEPL2015} we conclude that the qualitative 
physics of the non-linear model can be obtained on 
an $N = 4$ cluster.

\begin{figure}
 \includegraphics[width=0.7\columnwidth]{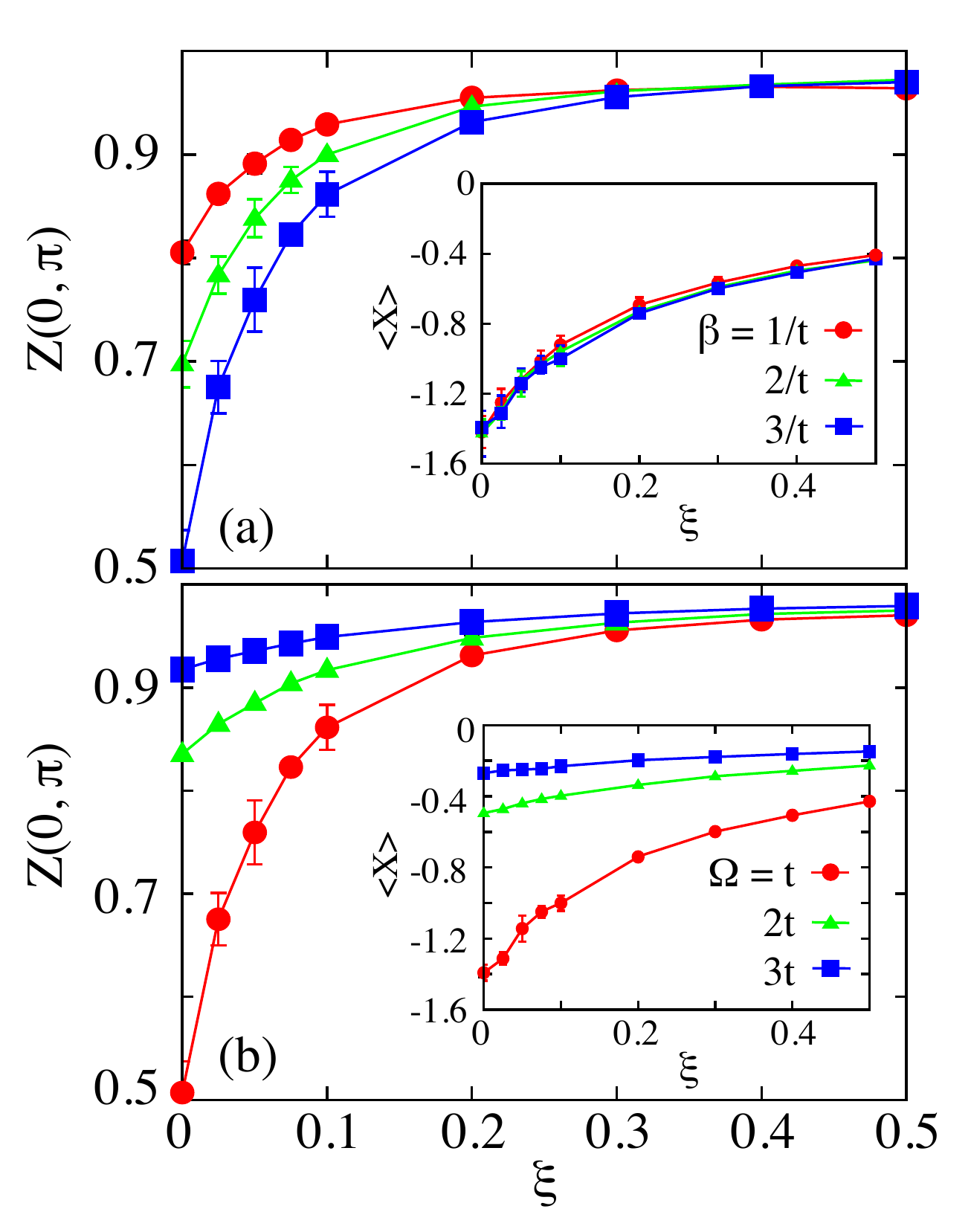}
 \caption{\label{Fig:Z_vs_beta} The (a) temperature and (b) $\Omega$ 
 dependence of the quasiparticle residue in the half-filled 
 model as a function of non-linear coupling strength $\xi$. 
 The insets show the corresponding expectation value of the lattice 
 displacement. All results are obtained on an $N = 4$ cluster 
 and with a linear coupling $\lambda = 0.25$. Error bars smaller than 
 the marker size have been suppressed for clarity. } 
\end{figure}

In Fig. \ref{Fig:Z_vs_beta}a we consider the temperature dependence 
of $Z(\bk)$ and average displacement of the lattice 
$\langle X \rangle = \frac{1}{N^2L}\sum_{i=1}^N\sum_{l=1}^L X_{i,l}$ for the 
half-filled model. Here, results for $Z(0,\pi)$ only are shown,   
since similar trends were found at all momenta. 
Focusing first on the linear model, we find that $Z(0,\pi)$ decreases 
with temperature as the CDW correlations begin to set in.  
The average lattice displacement, however, does not exhibit the same 
temperature dependence (see inset of Fig. \ref{Fig:Z_vs_beta}a).\cite{Footnote2}
As the non-linear interaction strength grows, however, the 
quasiparticle residue increases back towards its non-interacting value. 
For small $\xi$ this rise is somewhat 
rapid, but it gives way to a more gradual increase for $\xi \gtrsim 0.1$.  
The increase is also accompanied by a decrease in 
the average lattice displacement (inset of Fig. \ref{Fig:Z_vs_beta}). 
This behavior mirrors the observed $\xi$-dependence of the CDW 
susceptibility,\cite{LiEPL2015} and is consistent with the conclusion that 
a finite $\xi > 0$ undresses the carriers and relaxes the lattice distortions 
normally present in the linear model. 

The $\Omega$-dependence of $Z(\pi,0)$ and $\langle X \rangle$ for the 
same model are shown in  Fig. \ref{Fig:Z_vs_beta}b. 
Here, the $\xi = 0$ results are consistent with those obtained 
for the 1D Holstein model, where the tendency to form a CDW grows with decreasing 
phonon frequencies.\cite{Martin} Consequently, both  
the quasiparticle residue and average lattice displacement decrease as the 
value of $\Omega$ increases. The introduction of $\xi > 0$ results in the 
further decrease in these quantities.

From this section we conclude that the non-linear interaction with $\xi > 0$ 
acts to undress the quasiparticles and that this is a generic 
result, regardless of the values of $\Omega$ and $\beta$.  
The undressing, however, is much more pronounced at low temperatures, 
for smaller values of the phonon frequency, and near half-filling, 
where the CDW correlations (and subsequently the 
local lattice displacements) are largest in the linear model.

\subsection{Electron and Phonon energetics}  

\begin{figure}
 \includegraphics[width=\columnwidth]{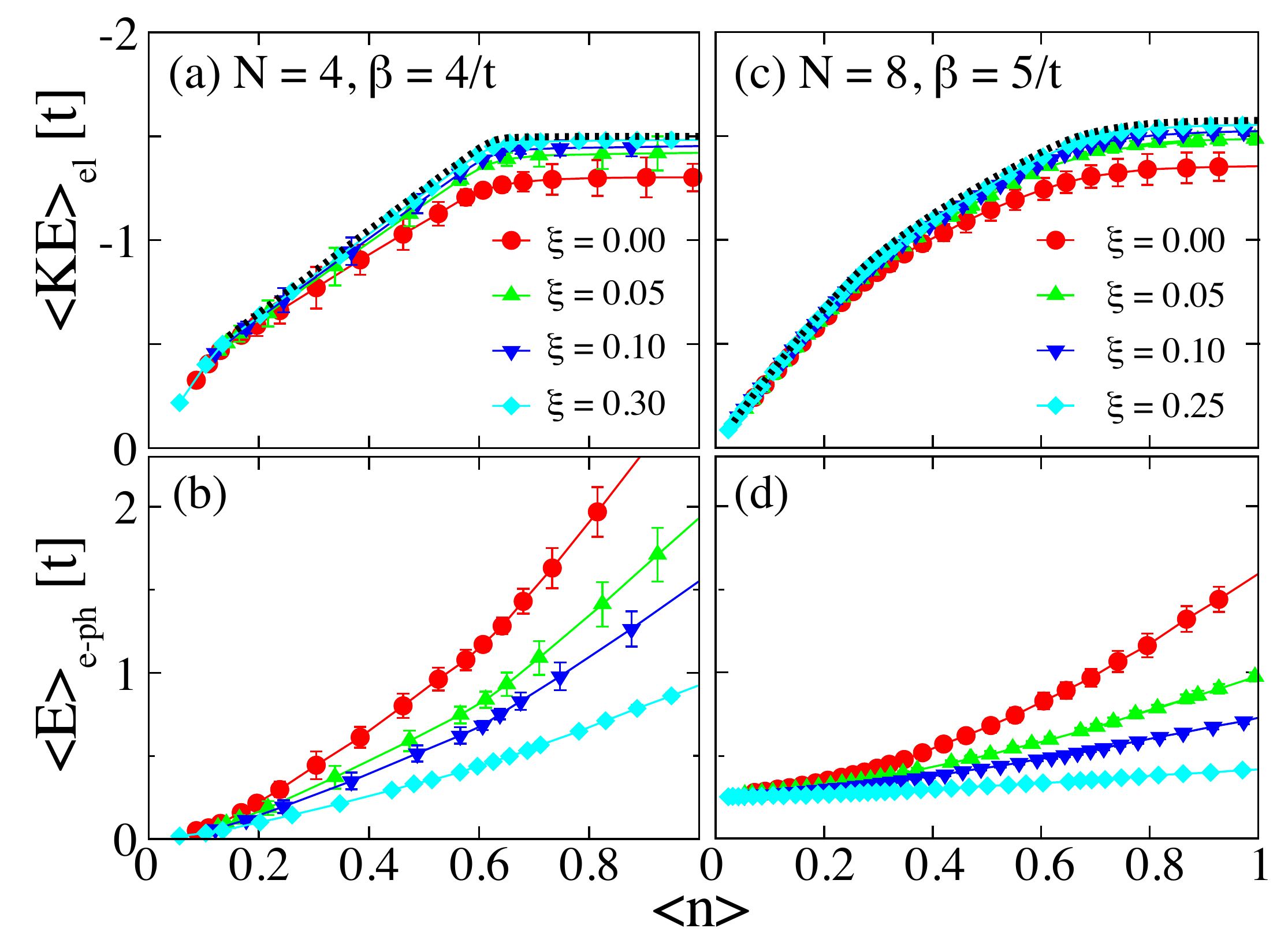}
 \caption{\label{Fig:elec_ener} (color online) 
 (a) The electron kinetic energy $\langle \text{KE}\rangle_\mathrm{el}$ and (b) the
 \ep energy $\langle\text{E}\rangle_\mathrm{e-ph}$ as a function of band filling $\langle
 n\rangle$ for various non-linear interaction strengths $\xi$, as
 indicated. Results are obtained on an $N=4$ cluster and with
 a linear coupling $\lambda=0.25$, phonon frequency $\Omega=t$, and an inverse 
 temperature $\beta=4/t$. The dashed lines in panels (a) and (c) indicate the 
 non-interacting result. (c) and (d) show corresponding results for a larger $N = 8$ 
 cluster and $\beta = 5/t$. The remaining parameters are the same as in (a) and (b). 
 Error bars smaller than the marker size have been
 suppressed for clarity.
}
\end{figure}

The average kinetic energy of the electronic subsystem 
$\langle \text{KE}\rangle_\mathrm{el}= -t\sum_{\langle i,j\rangle,\sigma}
\langle c_{i,\sigma}^{\dagger}c^\pdag_{j,\sigma}\rangle$ at $\beta = 4/t$ 
is shown in Fig. \ref{Fig:elec_ener}. 
Results are shown as a function of band filling $\langle n\rangle$ for various 
$\xi$ and for a linear coupling $\lambda = 0.25$. For $\xi=0$ 
the total kinetic energy $-\langle\text{KE}\rangle_e$ increases as a function 
of $\langle n\rangle$ as higher momentum states are populated in the Fermi sea,  
however, the total kinetic energy saturates as the 
filling increases beyond $\langle n\rangle > 0.6$. This is due to the 
saddle point in the band dispersion at $(0,\pi)$ and is 
also present in the non-interacting model (indicated by the dashed line).  
When the non-linear interaction is added we see an overall increase in the 
total kinetic energy, which tends towards the non-interacting 
value at all fillings for large $\xi$. This again reflects the undressing 
of the quasiparticles and the subsequent increase in mobility of the electronic subsystem. 

Fig. \ref{Fig:elec_ener}b shows the corresponding \ep interaction energy, 
defined as $\langle\text{E}\rangle_\mathrm{e-ph}=\sum_{i}\langle 
g_1\hat{n}_i\hat{X}_i+g_2\hat{n}_i\hat{X}_i^2\rangle$. Unsurprisingly, the total \ep interaction 
energy increases with band filling as both the average number of electrons 
per site and the average lattice displacement increase. This is 
most evident in the linear model ($\xi = 0$).  Increasing the 
value of $\xi$ naturally leads to smaller lattice displacements 
and a significant decrease in $\langle \text{E}\rangle_\mathrm{e-ph}$. 

\begin{figure}
 \includegraphics[width=\columnwidth]{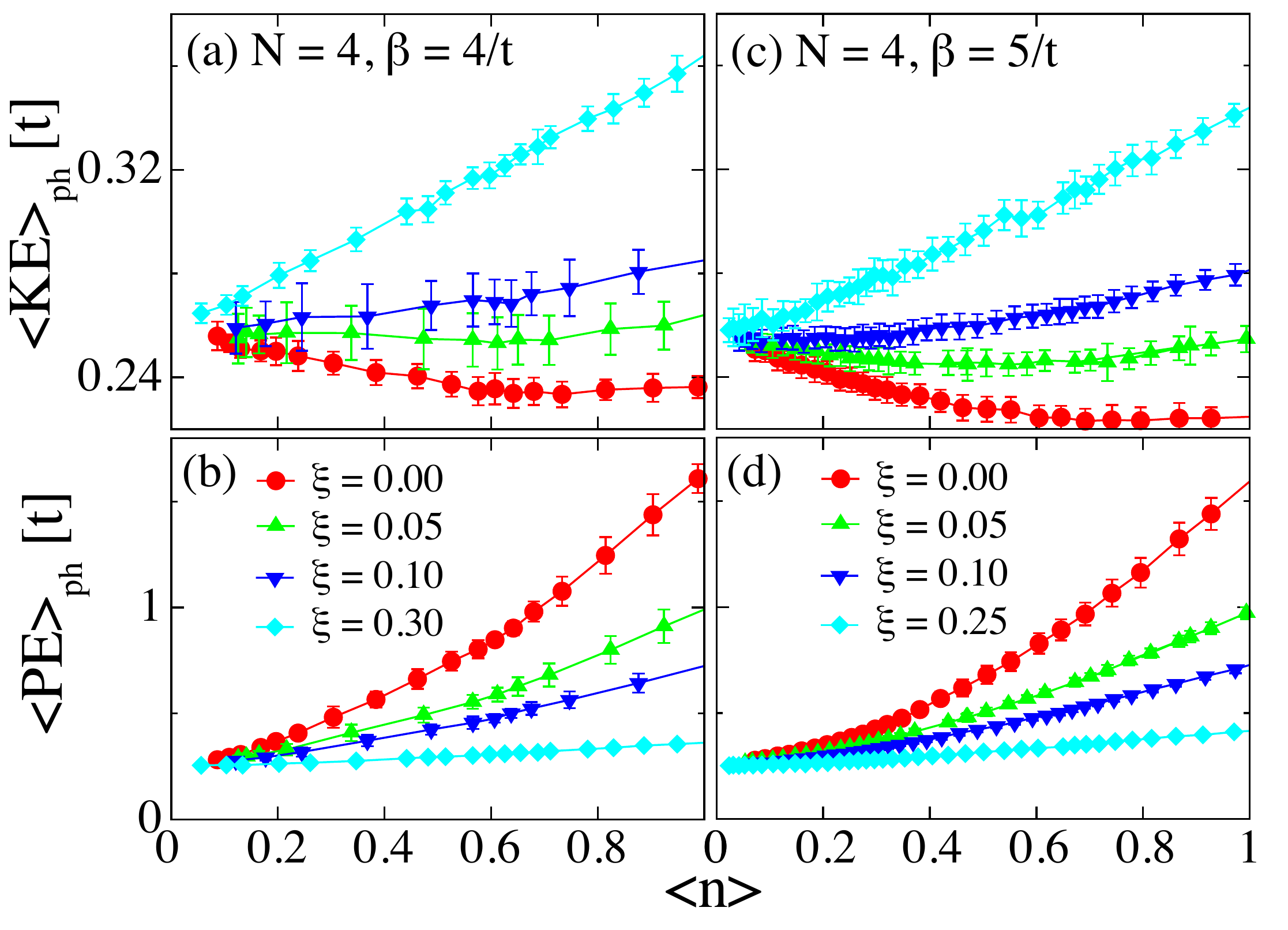}
 \caption{\label{Fig:phonon_ener} (color online)
 The phonon (a) kinetic energy $\langle \text{KE}\rangle_{ph}$ and (b) 
 potential energy $\langle PE\rangle_{ph}$ as a function of band filling
 $\langle n\rangle$ for various of the non-linear interaction strength $\xi$, as
 indicated in panel (b). Results are obtained on an $N=4$ cluster and with
 a linear coupling $\lambda=0.25$, $\Omega=1$,and $\beta=4/t$. (c) and (d) 
 show similar results obtained on a larger $N =8$ cluster with $\beta = 5/t$. 
 Error bars smaller than the marker size have been suppressed for clarity.
 }
\end{figure}

The average kinetic $\langle \text{KE} \rangle_\mathrm{ph}$ and 
potential $\langle \text{PE} \rangle_\mathrm{ph}$ 
energies of the lattice are shown in Figs. \ref{Fig:phonon_ener}a and 
\ref{Fig:phonon_ener}b, respectively. They are given by
\begin{eqnarray}
\langle \text{KE}\rangle_\mathrm{ph}&=&\frac{1}{2\Delta\tau}
-\frac{M}{2}\Big\langle\sum_{i,l}\left(\frac{X_{i,l+1}-X_{i,l}}{\Delta\tau}\right)^2\Big\rangle\\
\langle \text{PE}\rangle_\mathrm{ph}&=&\frac{M\Omega^2}{2}\Big\langle\sum_{i,l}X_{i,l}^2\Big\rangle. 
\end{eqnarray}
(The factor of $\frac{1}{2\Delta\tau}$ appearing in Eq. (10) is a
Euclidean correction introduced by the Wick rotation to the imaginary-time
axis.\cite{JohnstonPRB2013}) 

In the linear model we see a very weak variation in the phonon kinetic energy 
as a function of filling, with a slight decrease observed near half-filling when 
the CDW correlations increase. This is consistent with prior observations of the lattice 
kinetic energy in the vicinity of a CDW transition in the Hubbard-
Holstein model.\cite{JohnstonPRB2013} The 
average potential energy of the lattice grows as the average number of carriers per site 
increases. When the non-linearity is introduced and the 
lattice distortions diminish (see \ref{Fig:Z_vs_beta}b, inset), 
we see an increase in the lattice kinetic energy, which is attributed to the 
hardening of the phonon dispersion. At the same time, 
we see a decrease in the total lattice potential energy. 
Here, the non-linear interaction has  
two opposing effects: the increase in the phonon frequency increases the lattice potential 
energy while the decrease in the effective linear coupling decrease 
the net lattice distortions and subsequently lowers the potential energy. 
Our results indicate that the latter effect has the stronger impact.  

The energetics reported here are completely consistent with the conclusion that the 
non-linear interaction acts to harden the phonon frequency and weaken the effective linear 
interaction, which results in an undressing of the Holstein polaron for $\xi > 0$. 

\subsection{Phonon Spectral Properties}\label{Sec:AC} 
In the linear Holstein model the formation of the CDW phase is accompanied 
by a softening of the phonon dispersion to zero energy at the nesting 
wavevector ${\bf Q} = 2{\bf k}_F = (\pi,\pi)$. 
 \cite{VekicPRB1993, NowadnickPRB2015,WeberPreprint,Frank} This softening 
is generated by the strong nesting 
condition of the non-interacting Fermi surface. The inclusion 
of the non-linear \ep interaction is therefore expected to 
modify the phonon dispersion in two important ways: 
first, it will undo the 
softening at ${\bf Q}$ as the CDW correlations are suppressed. Second, it will 
result in an overall renormalization of the phonon frequency, as discussed in the 
introduction. We confirm these expectations in this section by examining the 
phonon spectral function $B({\bf q},\nu)$ and the phonon density of states (DOS) 
$N_{ph}(\nu) = \frac{1}{N^2}\sum_\bq B(\bq,\nu)$. 

\begin{figure}
 \includegraphics[width=0.75\columnwidth]{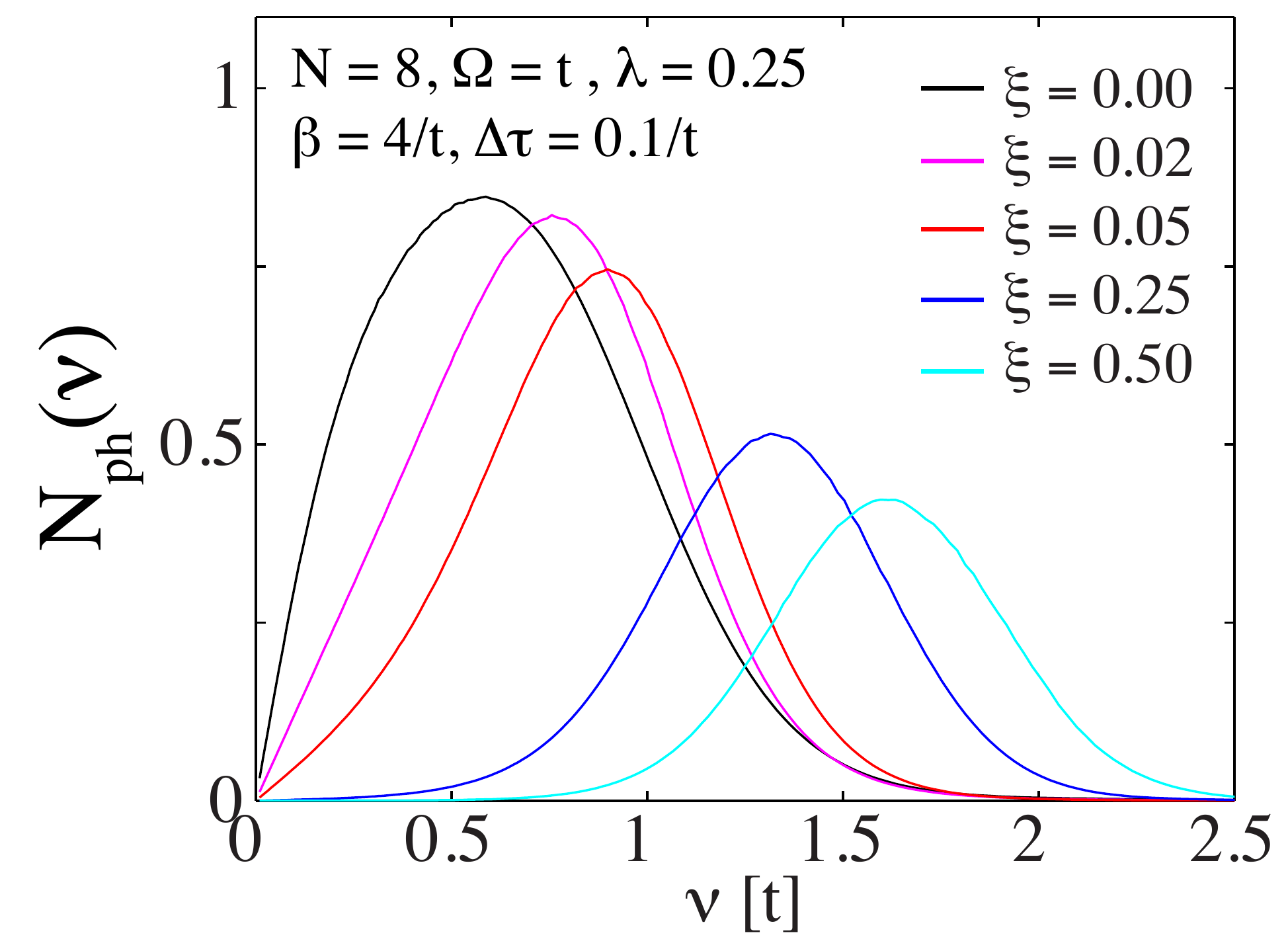}
 \caption{\label{Fig:PhononDOS} (color online) 
 The phonon density of states $N_\mathrm{ph}(\nu) = \frac{1}{N}\sum_\bq B(\bq,\nu)$ 
 for the half-filled model  
 as a function of the non-linear interaction strength. 
 The remaining calculation parameters are as indicated. 
 }
\end{figure}

\begin{figure}
 \includegraphics[width=0.9\columnwidth]{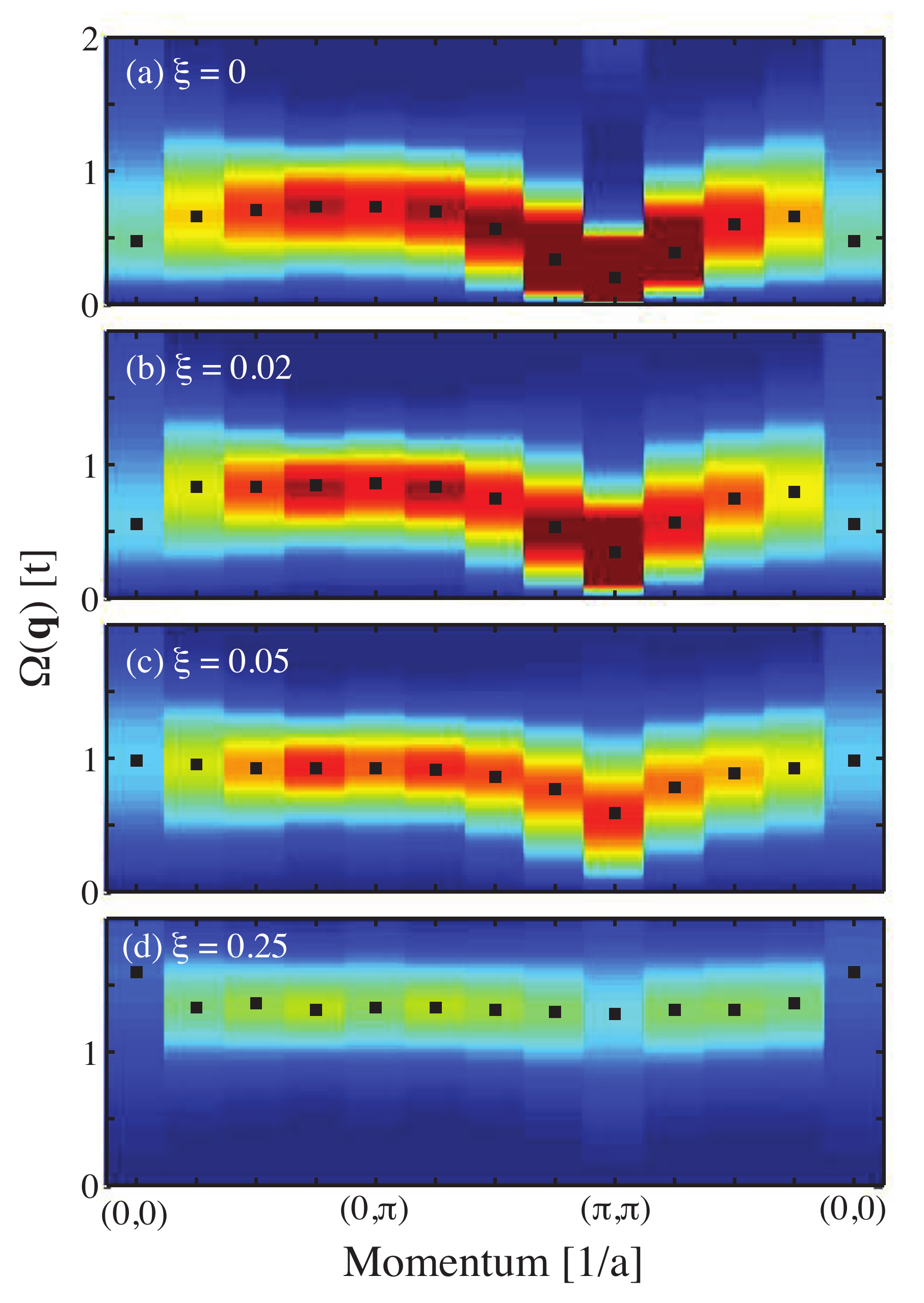}
 \caption{\label{Fig:omegaq} (color online) 
 The momentum resolved phonon spectral function $B(\bq,\nu)$ for the 
 half-filled model and for various values of the 
 non-linear interaction strength, as indicated in each panel. Results were obtained 
 on an $N = 8$ cluster with $\lambda = 0.25$, $\Omega = t$, $\beta = 4/t$ and 
 $\Delta\tau = 0.1/t$. The black squares indicate the position of the peak in the 
 phonon spectral function. 
 }
\end{figure}

In the non-interacting limit, $B(\bq,\nu)$ and  
$N_{ph}(\nu)$ are delta functions centered at the bare phonon frequency $\Omega$. 
In the presence of a non-zero linear interaction only, this distribution shifts to lower 
energy and broadens. This is illustrated 
in Figs. \ref{Fig:PhononDOS} and \ref{Fig:omegaq}a, which plot $N_{ph}(\nu)$ 
and the momentum-resolved phonon spectral function $B(\bq,\nu)$, respectively, 
for the half-filled model. These  
results were obtained on $N = 8$ clusters, with a linear coupling $\lambda = 0.25$, 
$\Omega = t$, and at an inverse temperature of $\beta = 4/t$.   
Due to the finite value of $\lambda$, the phonon frequency softens 
from its non-interacting value  
and $N_{ph}(\nu)$ for the linear model consists of a broad, asymmetric 
distribution centered at $\sim 0.60t$. 
The asymmetry in $N_{ph}(\nu)$ reflects the momentum dependence of the softening and the 
low-energy spectral weight in $B({\bf Q},\nu)$, coupled with the requirement 
that $B(0)=0$ for bosons. This  
is more easily seen in the momentum-resolved spectral function (Fig. \ref{Fig:omegaq}a), 
which has a clear Kohn anomaly at ${\bf Q} = (\pi,\pi)$. 

Two prominent changes occur when $\xi \ne 0$. 
First, the peak in the DOS shifts to 
higher energies, which verifies the hardening of the effective phonon 
frequency. This behavior is also clearly seen in the momentum resolved 
spectral functions, shown in Fig. \ref{Fig:omegaq}. 
Second, the pronounced Kohn anomaly begins to disappear  
as the CDW correlations are suppressed with increasing values of $\xi$.    
Both of these results confirm our expectations.  

\subsection{Mean-field Treatment of the quadratic \ep interaction}\label{Sec:MF} 
As we have repeatedly seen, the non-linear \ep interaction acts to   
renormalize both the bare linear interaction strength $\lambda$ 
and the bare phonon frequency $\Omega$. Both of these effects can be {\it qualitatively} 
understood at the mean-field (MF) level for the quadratic model, where an 
effective linear Hamiltonian is obtained by performing a MF decoupling of the 
interaction terms proportional to $b^\dagger_i b^\dagger_i$ and $b_ib_i$.\cite{AdolphsEPL2013} 
The resulting effective MF Hamiltonian is       
\begin{eqnarray}\label{Eq:MF} \nonumber
H_{MF}&=&H_{el}+\sum_i\Omega_{MF}\left(b_i^{\dagger}b_i^\pdag+\frac{1}{2}\right) \\ 
&+&\sum_{i,\sigma} g_{MF}\hat{n}_{i,\sigma}\left(b_i^{\dagger}+b_i^\pdag\right),
\end{eqnarray}
where $\Omega_{MF}=\Omega+2g_2$ and $g_{MF}=g_1(1-\frac{2g_2}{\Omega+4g_2})$ 
are the renormalized phonon frequency and \ep coupling constants, respectively. 
One immediately sees that the quadratic \ep interaction leads to a softening (hardening) 
of the phonon frequency and an increase (decrease) in the effective linear 
interaction strength $g_1$ for $\xi < 0$ ($\xi > 0$). These two effects 
combine to produce an overall increase (decrease) 
in the strength of the effective dimensionless coupling 
$\lambda_{eff} \propto \frac{g_{MF}^2}{\Omega_{MF}}$. 

\begin{figure}
 \includegraphics[width=0.8\columnwidth]{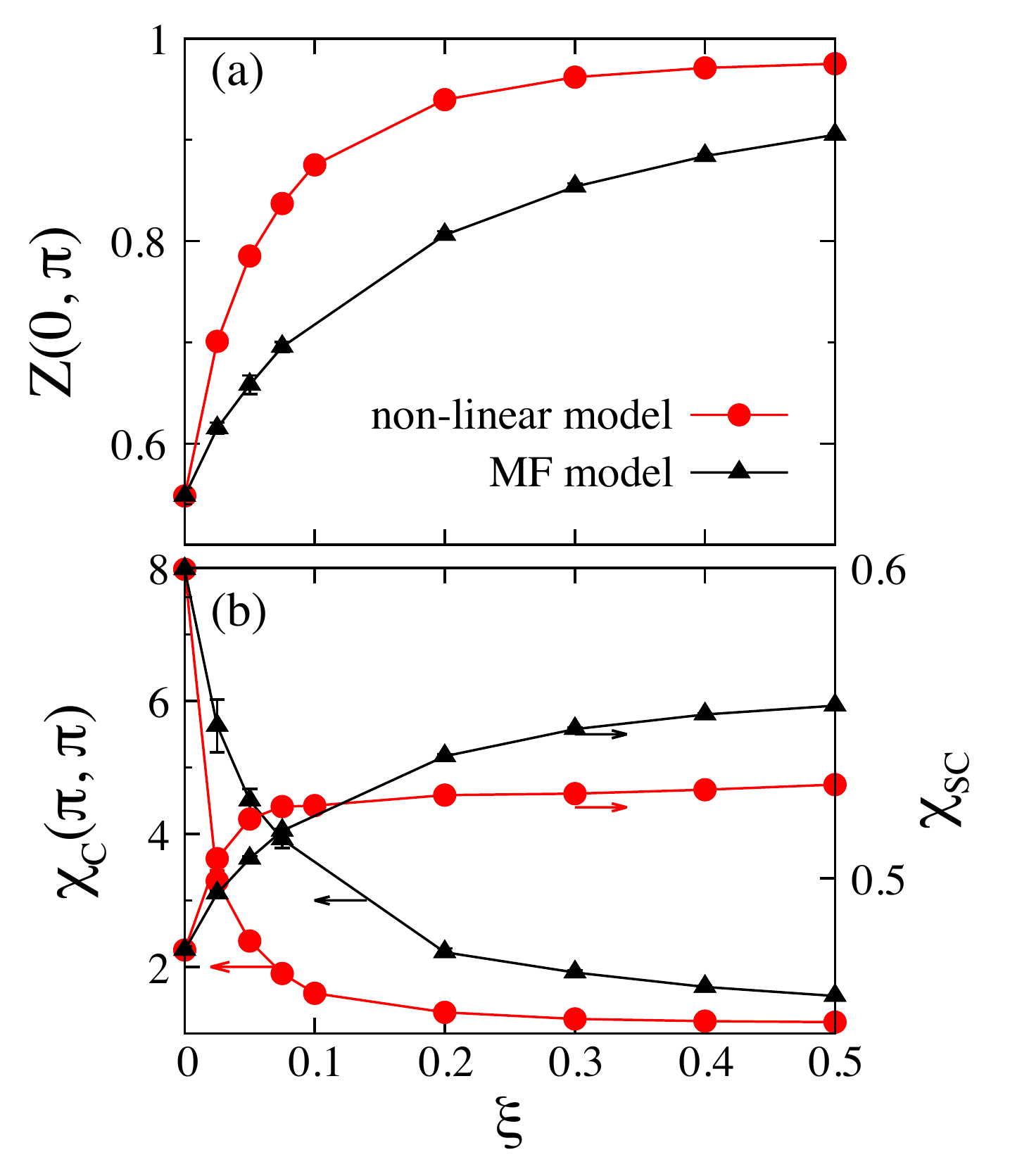}
 \caption{\label{Fig:Geff} (color online) A comparison of (a) 
 the quasiparticle residue and (b) CDW $\chi_C(\pi,\pi)$ and 
 pair field susceptibilities $\chi_{SC}$ 
 obtained from the non-linear model and its effective linear model, 
 as defined in the main text. 
 The bare linear coupling and phonon frequency are $\lambda = 0.25$ 
 and $\Omega = t$, respectively. 
 In both cases results are obtained on an $N = 8$ cluster and 
 at an inverse temperature of $\beta = 4/t$. Error bars smaller than 
 the marker size have been suppressed for clarity.} 
\end{figure}

\begin{figure}[t] 
 \includegraphics[width=\columnwidth]{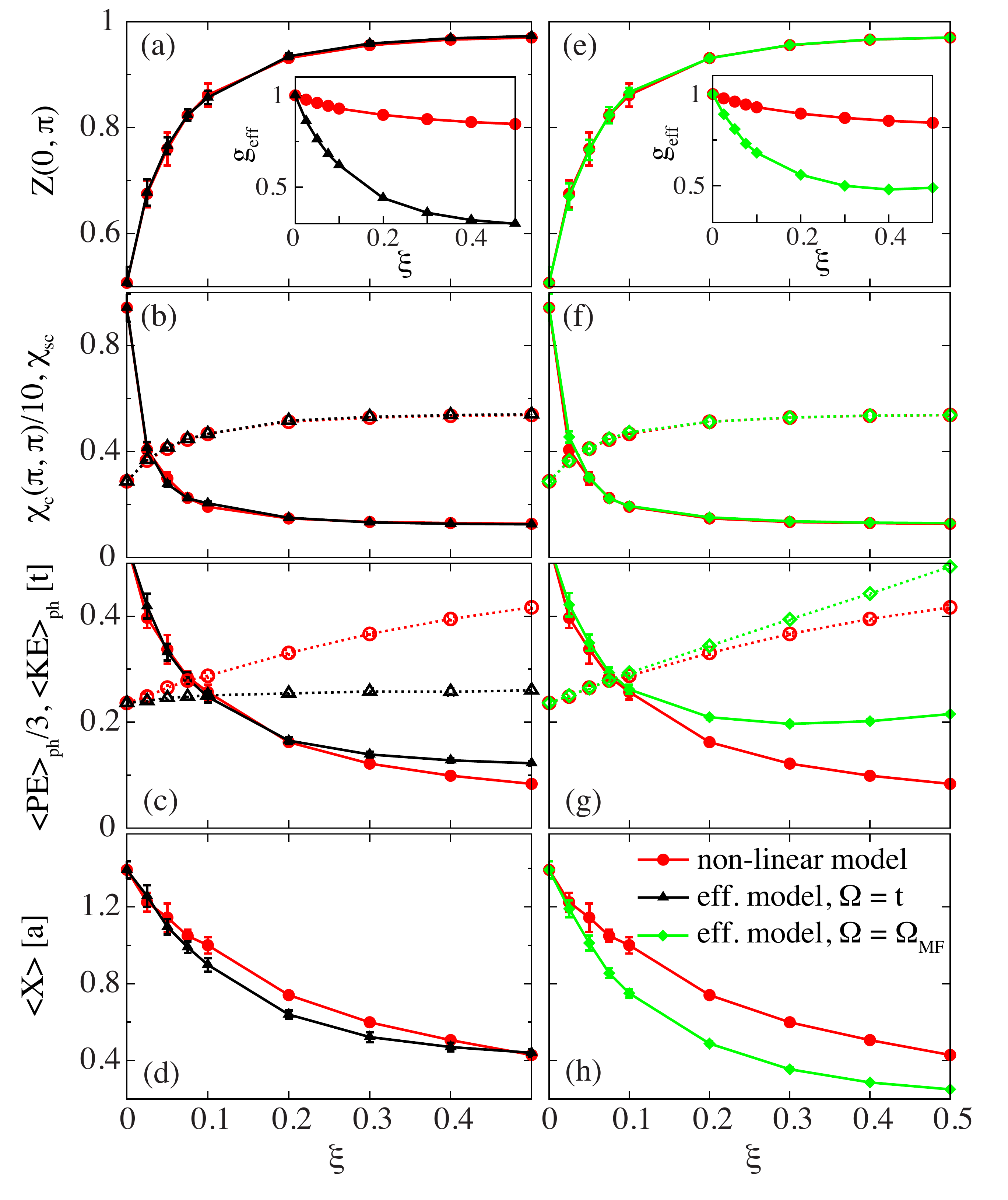}
 \caption{\label{Fig:Matching} (color online) A comparison of the results obtained 
 for the full non-linear model and an effective linear model where the value of the 
 \ep coupling constant has been adjusted to reproduce the electronic properties 
 of the non-linear model. Panels (a)-(d) show a comparison for an effective linear 
 model with $\Omega = t$, equal to the bare phonon frequency. Panels (e)-(f) 
 show a comparison for an effective linear model with $\Omega = \Omega_{MF}$. 
 The top row [panels (a) \& (e)] compares the quasiparticle residues obtained with 
 both models. The second row [panels (b) \& (f)] shows the resulting 
 charge (solid lines) and pair-field (dashed lines) susceptibilities. 
 The third row [panels (c) \& (g)] 
 show the resulting phonon potential and kinetic energies. The potential 
 energy has been divided by a factor of three and is indicated by the 
 solid lines while the kinetic energy is indicated by the dashed lines. Finally, the bottom 
 row [panels (d) \& (h)] show the average value of the lattice displacement. 
 The remaining parameters of the simulation are $\beta = 4/t$ and $N = 8$.}
\end{figure}

The MF treatment of the non-linear interaction is consistent with the general trends 
reported here and in Refs. \onlinecite{LiEPL2015} and \onlinecite{AdolphsEPL2013}. 
We stress, however, that the MF description only provides a qualitative 
picture of the non-linear effects. To illustrate this, we compare our DQMC
results for the full non-linear Hamiltonian against the predictions obtained from 
two sets of effective linear models. The
first is the MF-derived model defined by Eq. (\ref{Eq:MF}). The second 
is the set of effective linear models whose parameters are obtained by tuning 
the $\Omega_{eff}$ and $g_{eff}$ to reproduce the electronic 
properties of the system. 

We consider the MF-derived model first.  
Fig. \ref{Fig:Geff} compares the results for the quasiparticle residue, 
$\chi_{C}(\pi,\pi)$, and the pair-field susceptibility $\chi_{SC}$ calculated 
using the full non-linear model [Eq. (\ref{Eq:Full})] to results 
obtained from a DQMC simulation of the corresponding MF-derived linear model 
[Eq. (\ref{Eq:MF})] at half-filling.  
We find that the MF model does a poor job in quantitatively capturing 
the electronic properties; it underestimates both the quasiparticle residue  
$Z(0,\pi)$ and the tendency towards the formation of a CDW 
when compared to the full non-linear model. 
The MF model also over-predicts the magnitude of the 
superconducting pair-field susceptibility when $\xi$ is large and under-predicts 
it when $\xi$ is small.  
 
\begin{figure}
 \includegraphics[width=0.8\columnwidth]{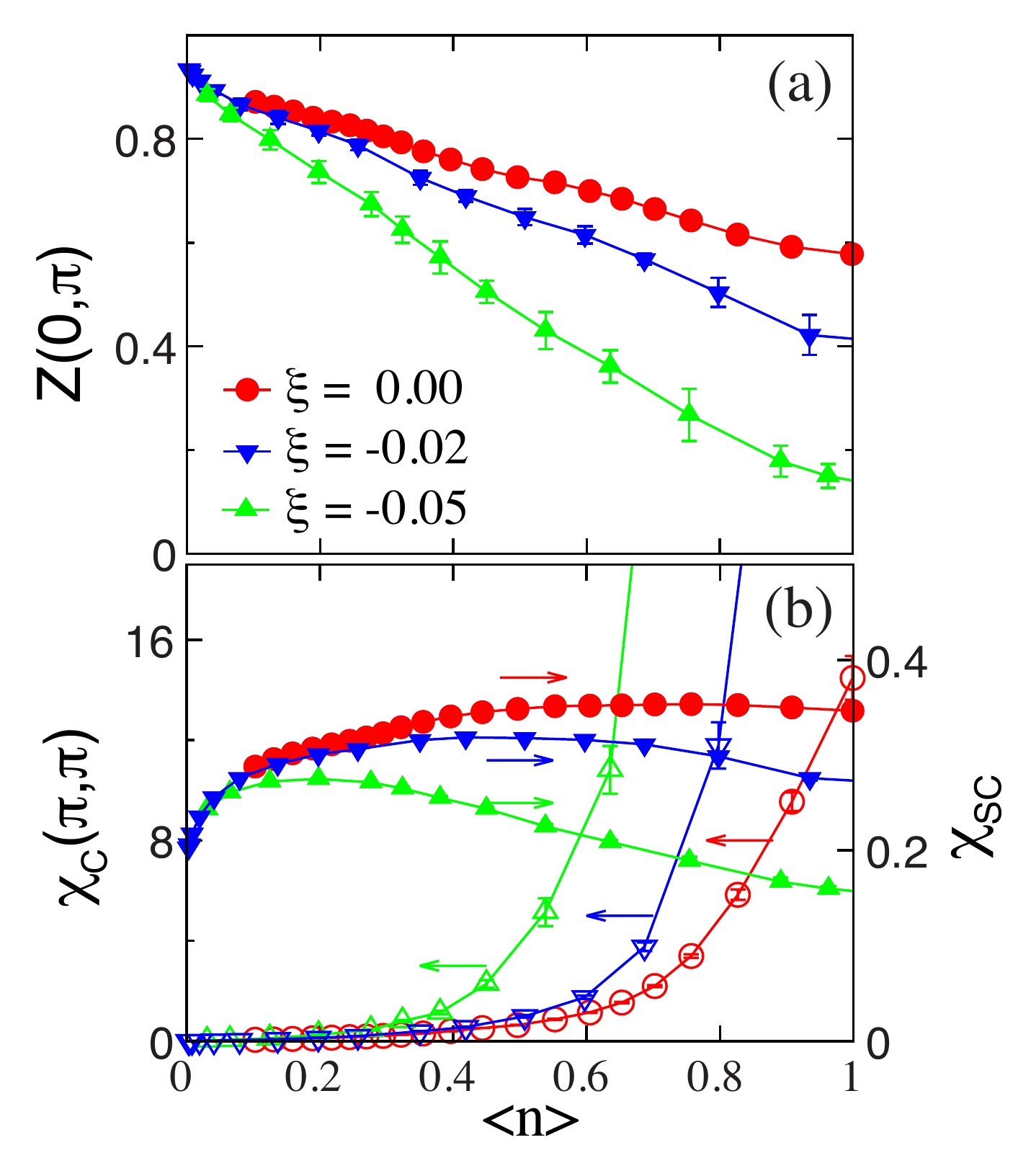}
 \caption{\label{Fig:Z_neg} (color online) (a) 
 the quasiparticle residue and (b) CDW $\chi_C(\pi,\pi)$ (open symbols) 
 and pair field susceptibilities $\chi_{SC}$ (solid symbols) 
 as a function of band filling.
 The  parameters are set as: $\lambda = 0.25$, $\Omega = 2t$, $\beta=4$. 
 The results are obtained on an $N = 8$ cluster. Error bars smaller than 
 the marker size have been suppressed for clarity.} 
\end{figure}

The results shown in Fig. \ref{Fig:Geff} demonstrate that 
the MF treatment of the quadratic interaction can only provide a qualitative  
picture of the physics of the non-linear model; however, another choice 
in effective model might do a better job. We explored this possibility 
by adjusting the effective coupling strength 
in the linear model such that the linear model reproduce 
the electronic properties of the full non-linear model. 
This procedure was performed for two choices in the phonon frequency. 
First, we set the phonon frequency equal to the bare value and adjusted the 
value of the coupling strength to reproduce the quasiparticle residue, as 
shown Fig. \ref{Fig:Matching}a. The value of the linear coupling strength 
$g_{eff}$ needed to produce this agreement is shown in the inset 
(black solid $\triangle$), where it is compared against the 
corresponding value of $g_{MF} = g_1 + 2g_2$. By tuning the value of 
$g_{eff}$ we are able to 
accurately capture the quasiparticle residue. The charge and 
superconducting pair-field susceptibilities are also well reproduced, indicating 
that this effective model is capable of capturing the electronic properties of 
the system. But when we examine the phonon properties (Figs. \ref{Fig:Matching}c 
\& \ref{Fig:Matching}d) we find some disagreement, 
particularity with respect to the predicted kinetic energy of the lattice, 
where the linear model systematically under-predicts the correct results. 

The comparison between the two models can be improved somewhat if we set the   
phonon frequency to be equal to $\Omega_{MF}$ and again readjust the 
value of $g_{eff}$. This case is shown in Fig. \ref{Fig:Matching}e-h. 
Using this choice we are again able to accurately capture the electronic 
properties and improve the comparison between the kinetic energy. 
But this comes at the expense of the level of agreement between the average lattice 
potential energy and the average lattice displacement. From this we conclude that 
an effective linear model cannot capture both the electronic and phononic properties 
of the non-linear model, for a fixed value of the phonon frequency.  
These results show that while the {\it qualitative effects} can be understood 
using an effective linear model, the full non-linear interaction should be 
retained if one wishes to accurately capture the effects of the non-linear 
interaction on both the phononic and electronic properties of the system.  
A similar conclusion was reached in Ref. \onlinecite{AdolphsEPL2013} in the
single carrier limit.

\begin{figure}
 \includegraphics[width=0.8\columnwidth]{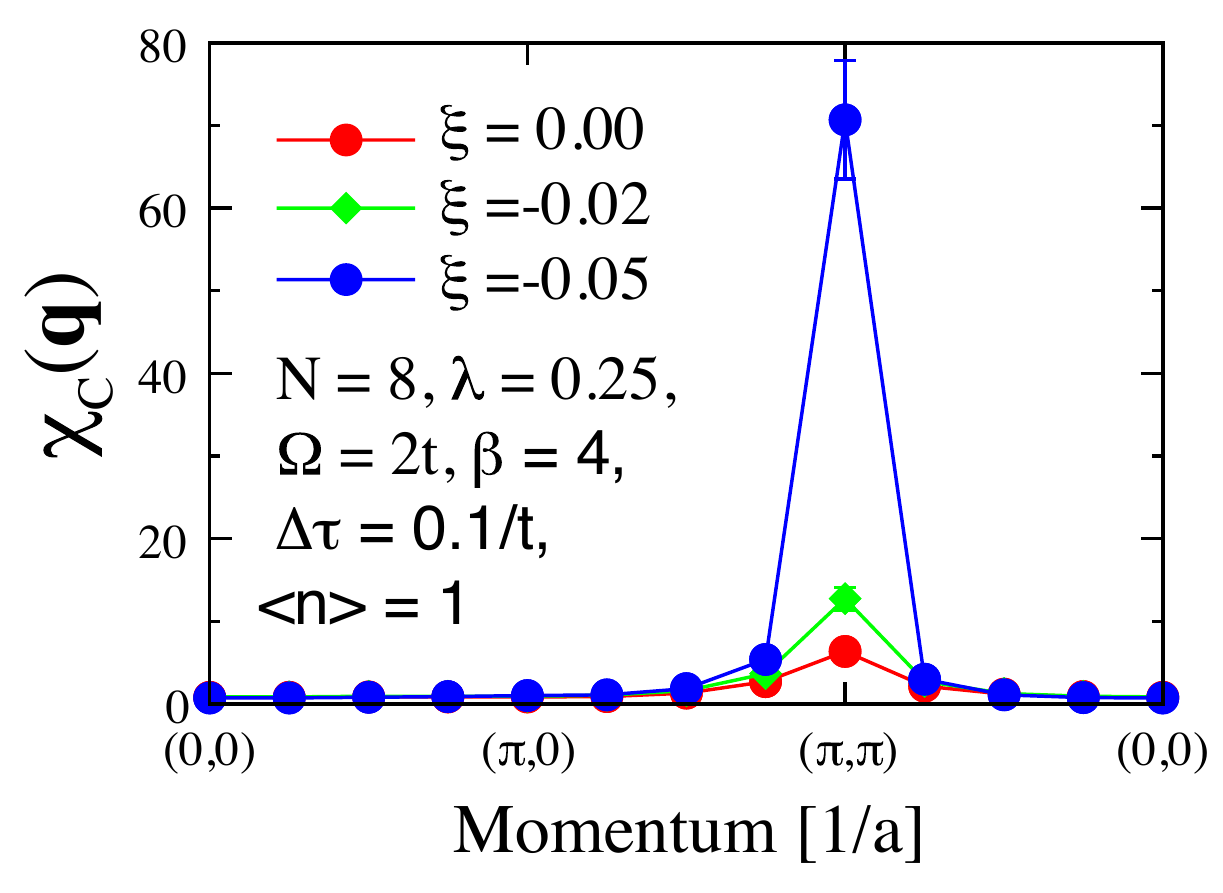}
 \caption{\label{Fig:Chiq_neg} (color online) The momentum dependence of the charge 
 susceptibility $\chi(\bq)$ as a function of non-linear interaction strength $\xi < 0$ at half filling. 
 The  parameters are set as: $\lambda = 0.25$, $\Omega = 2t$, $\beta=4$. 
 The results are obtained on an $N = 8$ cluster. 
 Error bars smaller than the marker size have been suppressed for clarity. 
 }
\end{figure}

\subsection{Negative values of $\xi$}\label{Sec:xineg}
We have shown that a positive ($\xi > 0$) non-linear coupling results in 
a hardening of the phonon frequency and a renormalization of the effective linear 
\ep coupling to weaker values. But what about the case when $\xi < 0$, where 
the MF model predicts an enhanced effective linear coupling? 
Before examining this case, we note that  
a large negative $\xi$ necessarily requires the inclusion of additional anharmonic 
terms in the lattice potential.\cite{AdolphsPRB2014} For $\xi < 0$ ($g_2 < 0$)
the phonon frequency given by $\Omega_{eff} = \Omega + 2g_2$ can become
negative for sufficiently large values of $g_2$, indicating an instability 
in the lattice. In this event the anharmonic terms of the lattice potential are 
required to maintain stability. At present, our codes do not 
contain such terms and we are unable to 
examine this case in great detail. We therefore restrict ourselves to a larger value of 
$\Omega = 2t$ and small values of $|g_2|$ in order to get a feel for the 
$g_2<0$ regime while ensuring the stability of the lattice.  

Fig. \ref{Fig:Z_neg}a shows the quasiparticle residue,  
$\chi_{C}(\pi,\pi)$, and $\chi_{SC}$ as a function of band filling for various
values of $\xi < 0$. These results were obtained for a linear coupling of 
$\lambda = 0.25$ and on an $N = 8$ cluster. We find that the quasiparticles  
are more effectively dressed when $\xi < 0$, and the quasiparticle residue 
is much smaller for all fillings when increasing negative quadratic interactions 
are included.  
The CDW correlations are also significantly enhanced, 
as reflected in the charge susceptibility shown in Fig. \ref{Fig:Z_neg}b. 
Both of these observations are in line with the expected increase in the 
effective linear coupling. Furthermore, since  
the CDW phase directly competes with s-wave superconductivity, the 
pair-field susceptibility is suppressed at filling values where the CDW 
correlations dominate. In addition, we also see a noticeable decrease in the 
pair-field susceptibility at band 
fillings where the CDW does not dominate. This suggests that the negative 
non-linear interaction  
has a detrimental effect on the superconducting transition temperature, which 
stems from the decrease in the quasiparticle residue. 
Fig. \ref{Fig:Chiq_neg} plots the momentum dependence of the charge 
 susceptibility $\chi(\bq)$ as a function of negative $\xi$ at half filling, 
where it is clear that the dominant CDW correlations are still being set by 
the Fermi surface nesting condition. 

\section{Summary and Conclusions}\label{Sec:conclude}
We have examined the role of non-linear \ep interactions in 
shaping the single-particle electronic and phononic 
properties of Holstein polarons at finite carrier concentrations 
and temperatures using DQMC. 
We find that the inclusion of a positive non-linear interaction term serves to
undress the polaron leading to carriers with lighter effective masses. This 
leads to changes in the energetics of both the electrons and phonons, as well as the 
relaxation of the local lattice distortions surrounding each carrier. 
This is due to a simultaneous hardening 
of the phonon frequency and renormalization of the effective linear coupling 
to smaller values. We have also examined the case when the quadratic \ep interaction 
has the opposite sign as the linear interaction, although this case cannot be 
explored in detail without the inclusion of additional anharmonic terms 
in the lattice potential. Nevertheless, in our limited range of accessible 
parameters, we find that a quadratic interaction 
results in an increased dressing of the carriers and an enhanced tendency towards 
the formation of a ${\bf Q} = (\pi,\pi)$ CDW ordered phase.  

While many of the effects we have discussed can be understood qualitatively 
at the mean-field level, we have demonstrated that the quantitative effects 
can only be captured by the full non-linear model. 
Specifically, the effective linear models fail to simultaneously 
capture the electronic and phononic properties. Therefore 
the full non-linear model must be retained if one wishes to accurately capture 
the properties of the electrons and the phonons. Our results are in good agreement 
with the results obtained in the single particle limit,\cite{AdolphsEPL2013} and show 
that non-linearities are relevant at finite carrier concentrations.  

\begin{acknowledgments}
We thank C. P. J. Adolphs and M. Berciu for useful discussions. 
This work is partially supported by the Science Alliance Joint Directed Research and 
Development (JDRD)  
program. CPU time for this work was provided in part by 
resources supported by the University of Tennessee and Oak Ridge National 
Laboratory Joint Institute for Computational Sciences (http://www.jics.utk.edu).
 
\end{acknowledgments}


\begin{thebibliography}{99}
\bibitem{Footnote} The dimensionless coupling 
$\xi$ is defined as $\xi = \frac{g_2}{g_1} = \left(\frac{\hbar}{2M_o\Omega}\right)^{\frac{1}{2}}\frac{\alpha_2}{\alpha_1}$.   
For a typical transition metal oxide  
$d_0 \sim 2$ $\AA$, $\beta = 3.5$, and $\left(\frac{\hbar}{2M_o\Omega}\right)^\frac{1}{2} \sim 0.043$ $\AA$ 
for an $\Omega = 70$ meV bond-stretching oxygen mode.
\bibitem{Footnote2}
This is due to the fact that the average lattice displacement is defined 
here as an average over all sites in the cluster. In the unordered uniform 
phase, each site is on average singly occupied. Conversely, in the 
CDW ordered phase, every second site is doubly 
occupied in a checkerboard pattern. If the average displacement 
of a singly occupied site is $X_0$, and the average displacement of a 
doubly occupied site is approximately twice this, then the average 
value of the displacement obtained by averaging over all sites  
will be $ \propto N^2X_0$ in the high-temperature unordered 
phase and $\propto \frac{1}{2}N^2(2X_0)$ in the low-temperature CDW phase. As a result, 
no significant temperate dependence is observed in the cluster 
averaged $\langle X \rangle$.
\end{thebibliography}
\end{document}